\documentclass[prd,showpacs,floatfix]{revtex4}

\usepackage{amsmath}
\usepackage{amsfonts}
\usepackage{amssymb}
\usepackage{bbm}
\usepackage[dvips]{graphicx}
\usepackage{color}

\begin{document}

\title{Transition redshift from the V-reconstruction method} 
\author{Wojciech Czaja}
\email{czaja@oa.uj.edu.pl}
\affiliation{Astronomical Observatory, Jagiellonian University,
Orla 171, 30-244 Krak\'ow, Poland}
\author{Marek Szyd{\l}owski}
\email{uoszydlo@cyf-kr.edu.pl}
\affiliation{Astronomical Observatory, Jagiellonian University, 
Orla 171, 30-244 Krak\'ow, Poland}
\author{Adam Krawiec}
\affiliation{Institute of Public Affairs, Jagiellonian University, 
Rynek G{\l}{\'o}wny 8, 31-042 Krak{\'o}w, Poland}
\email{uukrawie@cyf-kr.edu.pl}
\date{\today}

\begin{abstract}
We show the effectiveness of using the procedure of the potential reconstruction (the V-method) in the 
estimation of the transition redshift corresponding to switch from deceleration to acceleration phase of
the Universe. We investigate the FRW models with dark energy by applying the particle-like description of 
their dynamics. In this picture the evolution of the universe is represented by motion of a fictitious 
particle in the one-dimensional potential $V(a)$ or $V(z)$ where $a$ is the scale factor and $z$ is the 
redshift. The V-method solves the inverse problem where we reconstruct the potential function from empirical
data. We use the Riess et al.'s gold and gold+silver samples of SN Ia data in this reconstruction. In the 
same framework we obtain both the estimates of the form of the equation of state for dark energy and the 
transition redshift. We obtain that transition redshift is $0.38_{-0.07}^{+0.10}$ (gold subset) and 
$0.37_{-0.07}^{+0.06}$ (gold+silver sample) when the linear model of dark energy ($w(z)=w_{0} + w_{1}z$) and
$\Omega_{m,0}=0.3$ are assumed. We compare the estimation of transition redshift with the Riess et al. 
results. The V-method also alows us to find the value of Hubble function at the moment of transition.
\end{abstract}

\pacs{98.80.Bp, 98.80.Cq, 11.25.-w}

\maketitle

\section{Introduction}

The recent supernovae observations indicate that our universe is currently accelerating 
\cite{riess98,perlmutter99}. However, from the standard cosmological model we know that earlier the universe
decelerated because of domination of matter, and there should exist the moment when the switch from the 
deceleration phase to the acceleration phase takes place. The value of redshift for this transition $z_{T}$
has been recently found by Riess et al. \cite{riess04}. They obtained $z_{T} = 0.46 \pm 0.13$ using a 
kinematic model independent on the content of the universe. Their method does not allow to obtain both the 
equation of state parameters and the transition redshift in the same framework. The dark energy equation of 
state parameters are estimated in models with different ansatz on $w(z)$. However the constraint on the 
transition redshift was obtained using other relation for $d_{L}(z)$. Let us note that there is no reason 
to treat the value of $z_{T}$ estimated in a certain model as a real value in another model.

In this paper we propose the method which allows to calculate both equation of state parameters and the 
transition redshift in the same model. We estimate the value of $z_{T}$ through the potential function 
$V[a(z)]$ of the Hamiltonian system determining the evolution of the universe. In this approach the 
evolution of the universe is reduced to the motion of a fictitious particle with a unit mass in a one-dimensional potential. The information about the influence of matter content on the dynamics of the universe is 
derived from the shape of the potential function $V[a(z)]$. Namely the scale factor $a$ accelerates 
(decelerates) in the interval of $z$ if $V[a(z)]$ is an increasing (decreasing) function of $z$. Therefore 
we should expect the potential function has a maximum for the universe with deceleration during its earlier 
epoch and acceleration at present. 

In our previous papers \cite{szydlowski03,szydlowski03c} we developed the potential function method to probe
the dark energy. The use of the potential $V[a(z)]$ instead of $w_{X}(z)$ has an advantage because the
former suffers less from the smearing effect caused by the double integral which relates $w_{X}(z)$ and
$d_{L}(z)$ \cite{maor01}.
In these papers we showed that the dynamics of the Friedmann-Robertson-Walker (FRW) model of the universe, 
filled with non-interacting, non-relativistic matter with equation of state $p_{m}=w_{m}(a)\rho_{m}$ 
and dark energy with the equation of state $p_{X}=w_{X}(a)\rho_{X}$, can be reduced to the Hamiltonian 
system
\begin{equation}
\label{eq:3}
\mathcal{H} = \frac{\dot{a}^{2}}{2} + V(a),
\end{equation}
where an overdot denotes the differentiation with respect to the cosmic time $t$, and the potential 
\begin{equation}
\label{eq:4}
V(a) = - \frac{\rho_{\rm eff} a^{2}}{6}, \quad
\rho_{\rm eff}(a) = \rho_{m}(a) + \rho_{X}(a),
\end{equation}
where the units $8\pi G = c =1$ are used and $\rho_{\rm eff}$ is the effective energy density of the 
mixture of non-interacting fluids which satisfy the conservation equation 
\begin{equation}
\label{eq:5}
\dot{\rho}_{i} = -3 \frac{\dot{a}}{a} (\rho_{i} + p_{i}).
\end{equation}
Then $\rho_{\rm eff}$ also satisfies equation (\ref{eq:5}).
On the other hand the potential function $V(a)$ can be reconstructed from the SN Ia data through the 
Hubble function $H(z)$ which in the spatially flat universe is related to the luminosity distance 
$d_{L}(z)$ as
\begin{equation}
\label{eq:2}
H(z) = \left[\frac{d}{dz} \left( \frac{d_{L}(z)}{1+z} \right)\right]^{-1}. 
\end{equation}

Because the trajectories of system (\ref{eq:3}) always lie on the zero-energy level (as a consequence of
the Hamiltonian constraint) we can rewrite potential (\ref{eq:4}) to the form
\begin{equation}
\label{eq:6}
V(z) = - \frac{1}{2} \left[ (1+z) \frac{d}{dz} \left( \frac{d_{L}(z)}{1+z} \right) \right]^{-2}.
\end{equation}

The potential $V(z)$ contains all information necessary to determine the full dynamics of the universe on 
the phase plane $(a,\dot{a})$. Thus both the effective energy density $\rho_{\rm eff}(a)$ and the equation 
of state coefficient $w_{\rm eff}[a(z)] \equiv p_{\rm eff}/\rho_{\rm eff}$ can be calculated unambiguously 
from the potential function $V(a)$ 
\begin{equation}
\label{eq:7}
\rho_{\rm eff}(a) = - \frac{6V(a)}{a^{2}}, \quad w_{\rm eff}(a) = - \frac{1}{3} [1 + I_{V}(a)],
\end{equation}
where $I_{V}(a) = (d \ln{V})/(d \ln{a})$ is the elasticity of the potential function $V$ with respect to the
scale factor $a$. 

As it is well known the behavior of the potential function in the neighborhood of a maximum can be 
approximated by a quadratic part of the expansion in Taylor's series. Therefore we have 
\begin{equation}
\label{eq:8}
V(a) = V(a_{T}) + \frac{1}{2} 
\left. \frac{\partial^{2} V}{\partial a^{2}}
\right|_{a=a_{T}} (a-a_{T})^{2}
\end{equation}
where $V(a_{T}) = -\dot{a}_{T}^{2}/2$ from the Hamiltonian constraint, and
\begin{equation}
\label{eq:9}
\rho_{\rm eff}(a) = - \frac{6}{a^{2}} \left[ 
-\frac{\dot{a}_{T}^{2}}{2} + 
\frac{1}{2} \left. \frac{\partial^{2} V}{\partial a^{2}}
\right|_{a=a_{T}} (a-a_{T})^{2} \right]
\end{equation}
or  
\begin{align}
\rho_{\rm eff}(z) =& - 3 (1+ z_{T})^{4} \left. 
\frac{\partial^{2} V}{\partial z^{2}} \right|_{z=z_{T}} \nonumber \\
& + 3 (1+z)^{2} \left[ \frac{1}{(1+z_{T})^4} 
\left( \left. \frac{d z}{d t} \right|_{z=z_{T}} \right)^{2}
- (1+z_{T})^{2} \left. \frac{\partial^{2} V}{\partial z^{2}}
\right|_{z=z_{T}} \right] \nonumber \\
& - 6(1+z) (1+z_{T})^{3} \left. \frac{\partial^{2} V}{\partial z^{2}} 
\right|_{z=z_{T}}.
\label{eq:10} 
\end{align}

In the case of transition from deceleration to acceleration one can see from formula~(\ref{eq:10}) that 
there are three terms which can be interpreted as the cosmological constant (positive), the curvature 
contribution (negative), and the topological deffects with positive energy ($p=-2/3 \rho$). 
Note that as $z$ is close to $-1$ ($a \gg a_{0}$) than the last two terms are negligible and only the 
``cosmological constant'' dominates the dynamics of the late epoch of the universe. 

The value of the potential function at the maximum is given by 
\begin{equation}
\label{eq:11}
V(a_{T}) = -\frac{1}{2} H_{T}^{2}(1+z_{T})^{-2}.
\end{equation}
The value of $z_{T}$ and $V(z_{T})$ inform us about the moment of transition and the Hubble function
at this moment. The estimation of these parameters will be done in the next section.

\section{Estimation of redshift transition}

In our previous papers the method of reconstruction of the potential function was applied to the analysis of 
global dynamics of the FRW models in the phase space \cite{szydlowski03,szydlowski03c,szydlowski04}. 
The potential function was reconstructed from the SN Ia data which allow us to determine all evolutional 
paths of the model for all admissible initial conditions. These results have the qualitative character, but 
it is also possible to obtain some quantitative attributes of the reconstructed dynamics of a model. 

It seems to be important to determine when the switch from the deceleration to the acceleration happens in 
different cosmological models. Because it is possible to differentiate between cosmological models through 
the quantitative features of their potential functions, we propose to use it as the test of cosmological 
models. The maximum of the potential function, interpreted as the time of switch between deceleration and 
acceleration phases, can be chosen as a one of the criteria (another may be the value of the potential 
function at $z_{T}$). If someone obtains the value of $z_{T}$ independently from any model 
(e.g. one can numerically probe it directly from the observational data), it could be compared with values 
of $z_{T}$ found in different cosmological models. 

We calculate the probability distribution of the transition redshift for the spatially flat FRW model filled 
with nonrelativistic matter ($\rho_{m}=\rho_{m,0}a^{-3}$) and dark energy for which the linear 
parametrization of the equation of state is assumed ($w(z) = w_{0} + w_{1} z$). In our analysis we use the 
recently released SN Ia data by Riess et al. \cite{riess04}. We use both the gold subset and gold+silver 
(full) sample. To estimate the cosmological model parameters $w_{0}$, $w_{1}$, and the transition redshift 
$z_{T}$ ($\Omega_{m,0}=0.3$ and $H_{0}=65 {\rm km}$ ${\rm s}^{-1}$ ${\rm Mpc}^{-1}$ priors are used) we 
minimize the $\chi^{2}$ statistics given by the formula
\begin{equation}
\label{eq:12}
\chi^{2}(\Omega_{m,0},H_{0},w_{0},w_{1})=\sum_{j}\frac{\left[\mu_{t,j}(z_{j};\Omega_{m,0},H_{0},w_{0},w_{1})-
\mu_{o,j}\right]^{2}}{\sigma_{\mu_{o,j}}^{2}},
\end{equation}
where $\mu_{o,j}=m_{o,j}-M$ is the extinction-corrected distance modulus for SN Ia at redshift $z_{j}$, 
$\sigma_{\mu_{o,j}}$ is the uncertainty in the distance moduli $\mu_{o,j}$ including the dispersion
in galaxy redhift due to peculiar velocities, 
$\mu_{t,j}(z_{j};\Omega_{m,0},H_{0},w_{0},w_{1})=5\log{d_{L}(z_{j};\Omega_{m,0},H_{0},w_{0},w_{1})}+25$,
\begin{gather}
d_{L}(z_{j};\Omega_{m,0},H_{0},w_{0},w_{1})
=(1+z_{j})\int_{0}^{z_{j}}\frac{dz'}{H(z';\Omega_{m,0},H_{0},w_{0},w_{1})} \nonumber \\ 
=(1+z_{j})H_{0}^{-1}\int_{0}^{z_{j}}\frac{dz'}{\sqrt{\Omega_{m,0}(1+z')^{3}+\Omega_{X,0}(1+z')^{3(1+w_{0}-w_{1})}\exp{(3w_{1}z)}}}.
\label{eq:13}
\end{gather}
The sumation in equation (\ref{eq:12}) is over all of the observed supernovae. 
The potential function for the model assumes the form
\begin{gather}
V[a(z),\Omega_{m,0},H_{0},w_{0},w_{1}]=-\frac{1}{2}H^{2}(a(z),\Omega_{m,0},H_{0},w_{0},w_{1})a^{2}(z)
=-\frac{1}{2}\left[\frac{d}{dz}\left(\frac{d_{L}(a(z),\Omega_{m,0},H_{0},w_{0},w_{1})}{1+z}\right)\right]^{-2}a^{2}(z) \nonumber \\
=-\frac{1}{2}H_{0}^{2}\left[\Omega_{m,0}(1+z)+\Omega_{X,0}(1+z)^{3(1+w_{0}-w_{1})-2}\exp{(3w_{1}z)}\right].
\label{eq:14}
\end{gather}

\begin{table}[!ht]
\caption{Fitting results of the statistical analysis from distant type Ia supernovae data.\\
(Sample of 186 SN Ia, Riess et al. (2004))}
\begin{center}
\begin{ruledtabular}
\begin{tabular}{c|c|c|c|c|c|c|c|c}
Sample&$\Omega_{m,0}$&$\Omega_{X,0}$&$w_{0}$ & $w_{1}$&$\chi^{2}$&$z_{T}$&$V(z_{T})$&Method\\
\hline
gold		& $0.30$ & $0.70$ & $-1.35$ & $1.52$ & $175.57$ & $0.40$ & $-1.60$ & best fit \\
& $0.30$ & $0.70$ & $-1.33^{+0.19}_{-0.19}$ & $1.49^{+0.81}_{-0.94}$ && $0.38^{+0.10}_{-0.07}$ &$-1.58^{+0.05}_{-0.06}$ & max(P)\\
\hline
gold + silver	& $0.30$ & $0.70$ & $-1.41$ & $1.73$ & $229.36$ & $0.38$ & $-1.58$ & best fit \\
& $0.30$ & $0.70$ & $-1.39^{+0.17}_{-0.18}$ & $1.70^{+0.75}_{-0.85}$ && $0.37^{+0.07}_{-0.06}$ &$-1.58^{+0.05}_{-0.04}$ & max(P)\\
\end{tabular}
\end{ruledtabular}
\end{center}
\label{results1}
\end{table}
  
\begin{figure}[!ht]
\begin{center}
$\begin{array}{c@{\hspace{0.2in}}c}
\multicolumn{1}{l}{\mbox{\bf (a)}} & 
\multicolumn{1}{l}{\mbox{\bf (b)}} \\ [0.0cm]
\includegraphics[scale=0.4, angle=0]{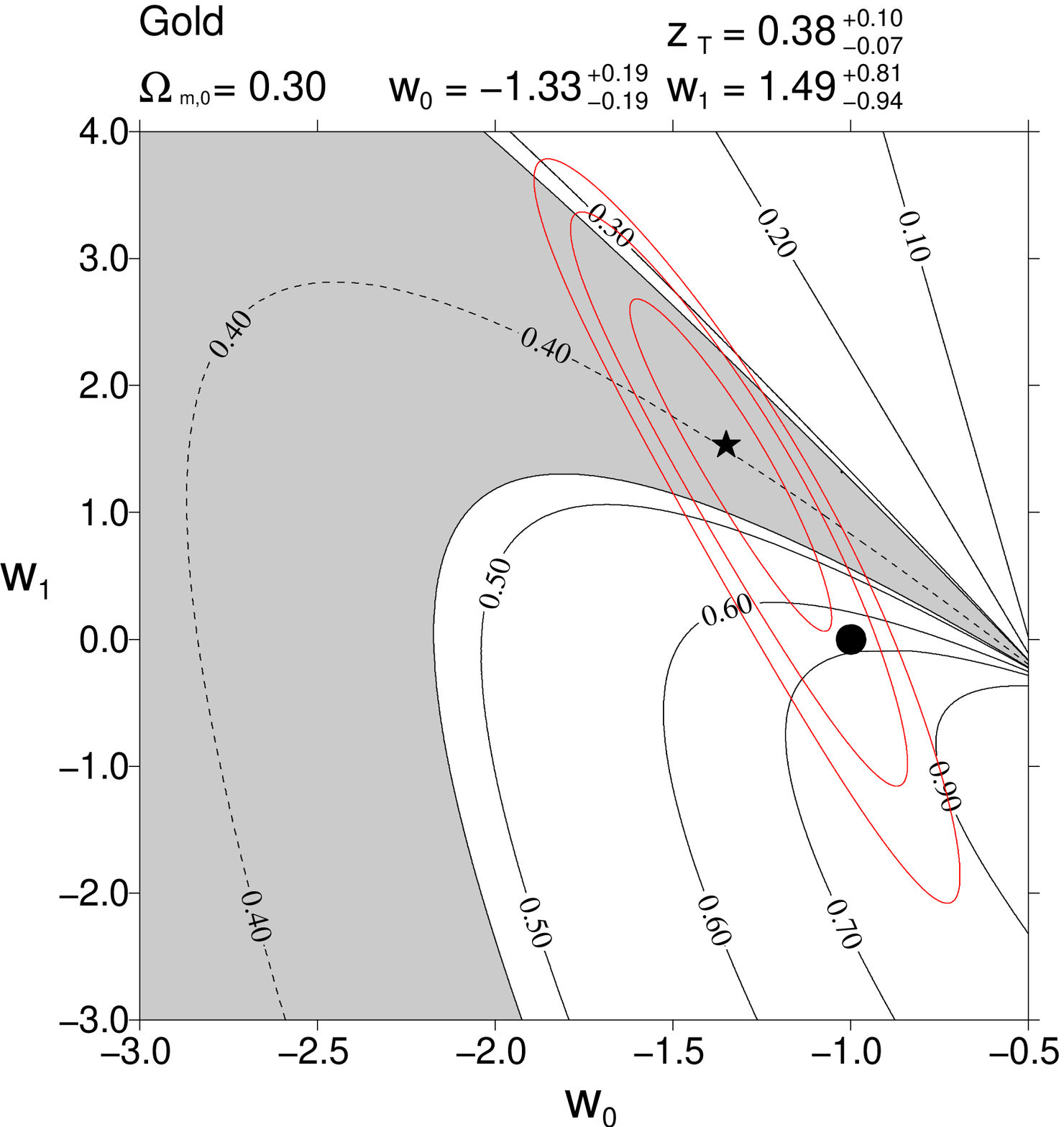} & 
\includegraphics[scale=0.4, angle=0]{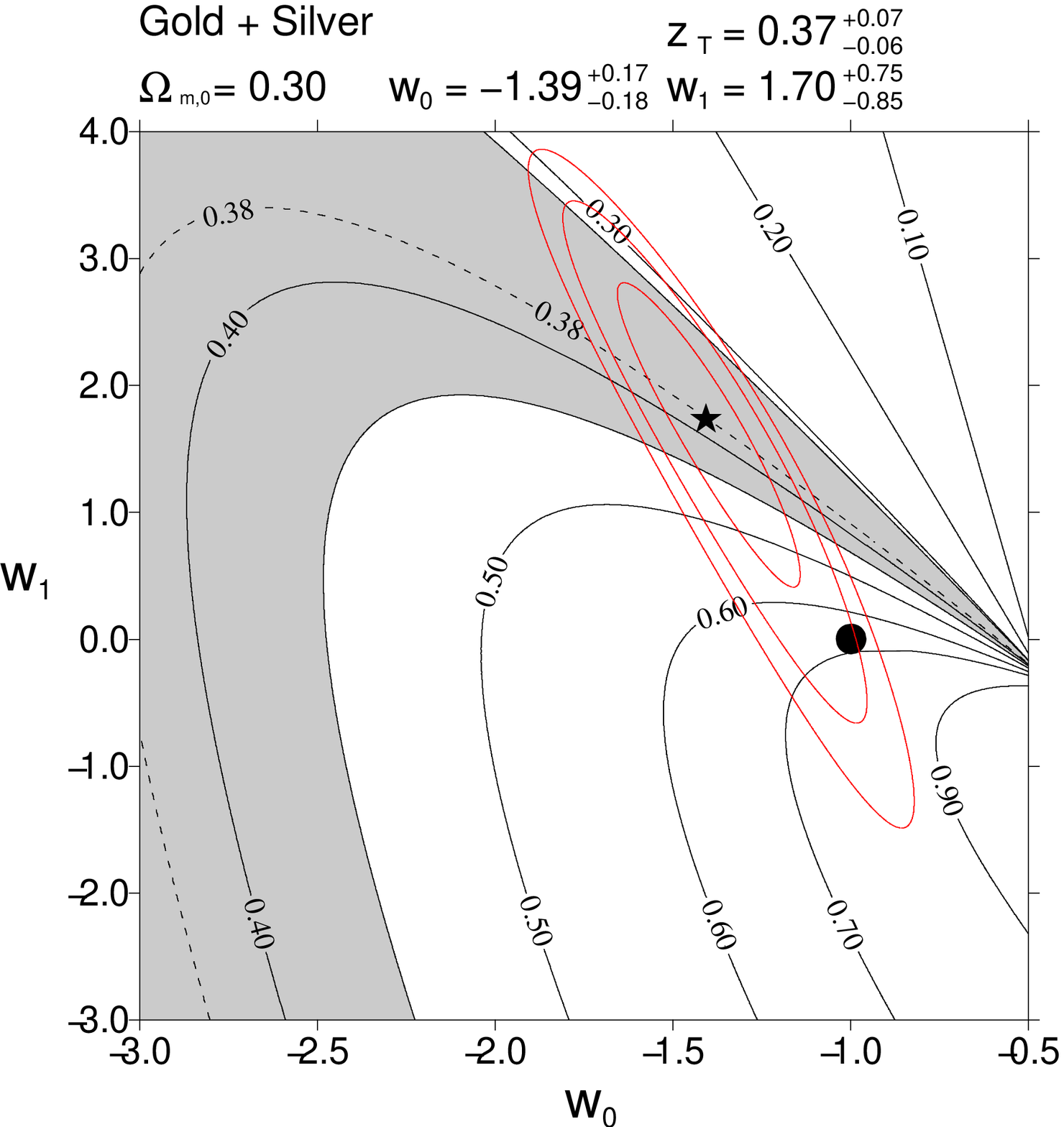} \\ [0.4cm]
\multicolumn{1}{l}{\mbox{\bf (c)}} & 
\multicolumn{1}{l}{\mbox{\bf (d)}} \\ [-0.5cm]
\includegraphics[scale=0.3, angle=270]{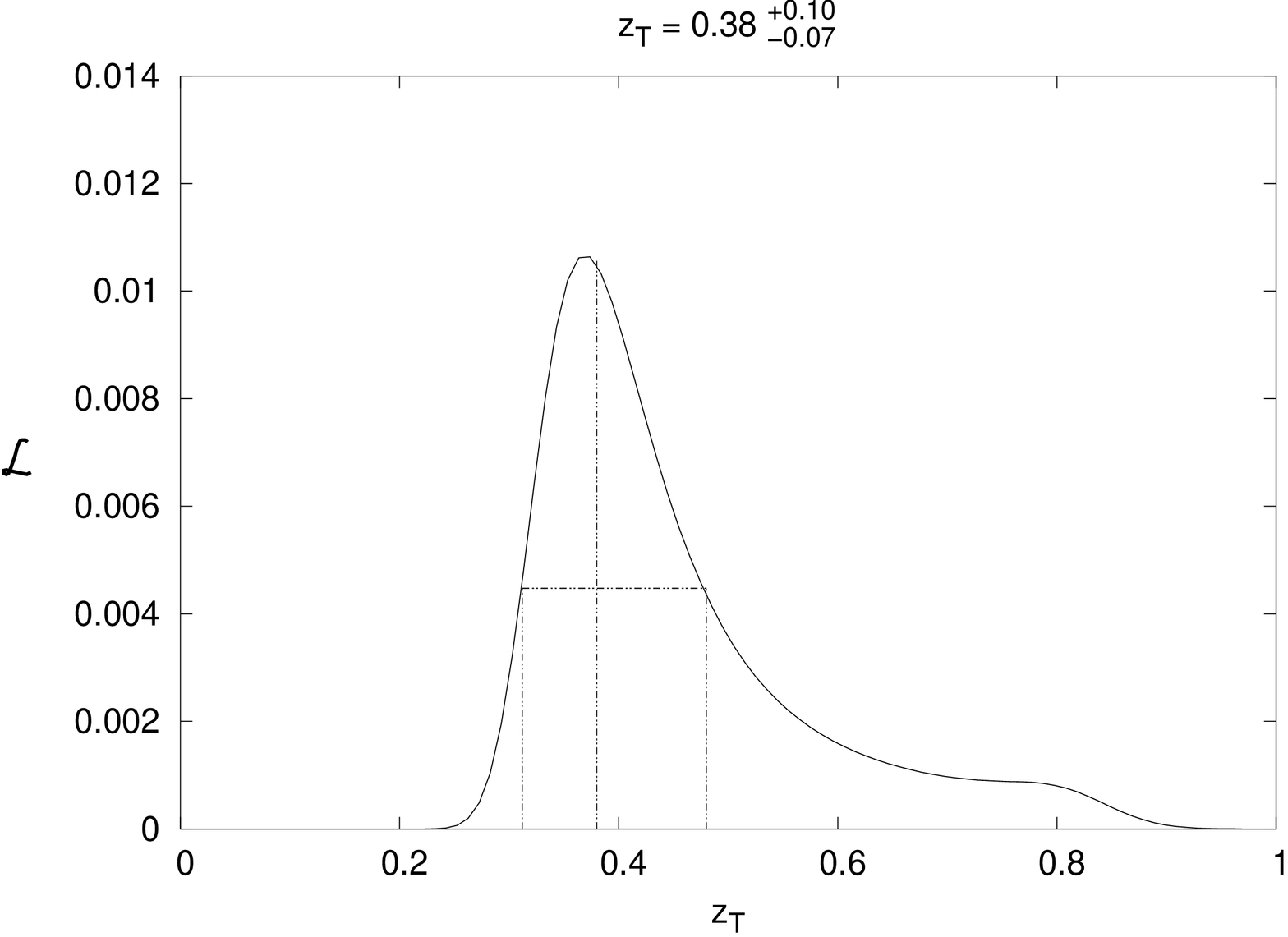} & 
\includegraphics[scale=0.3, angle=270]{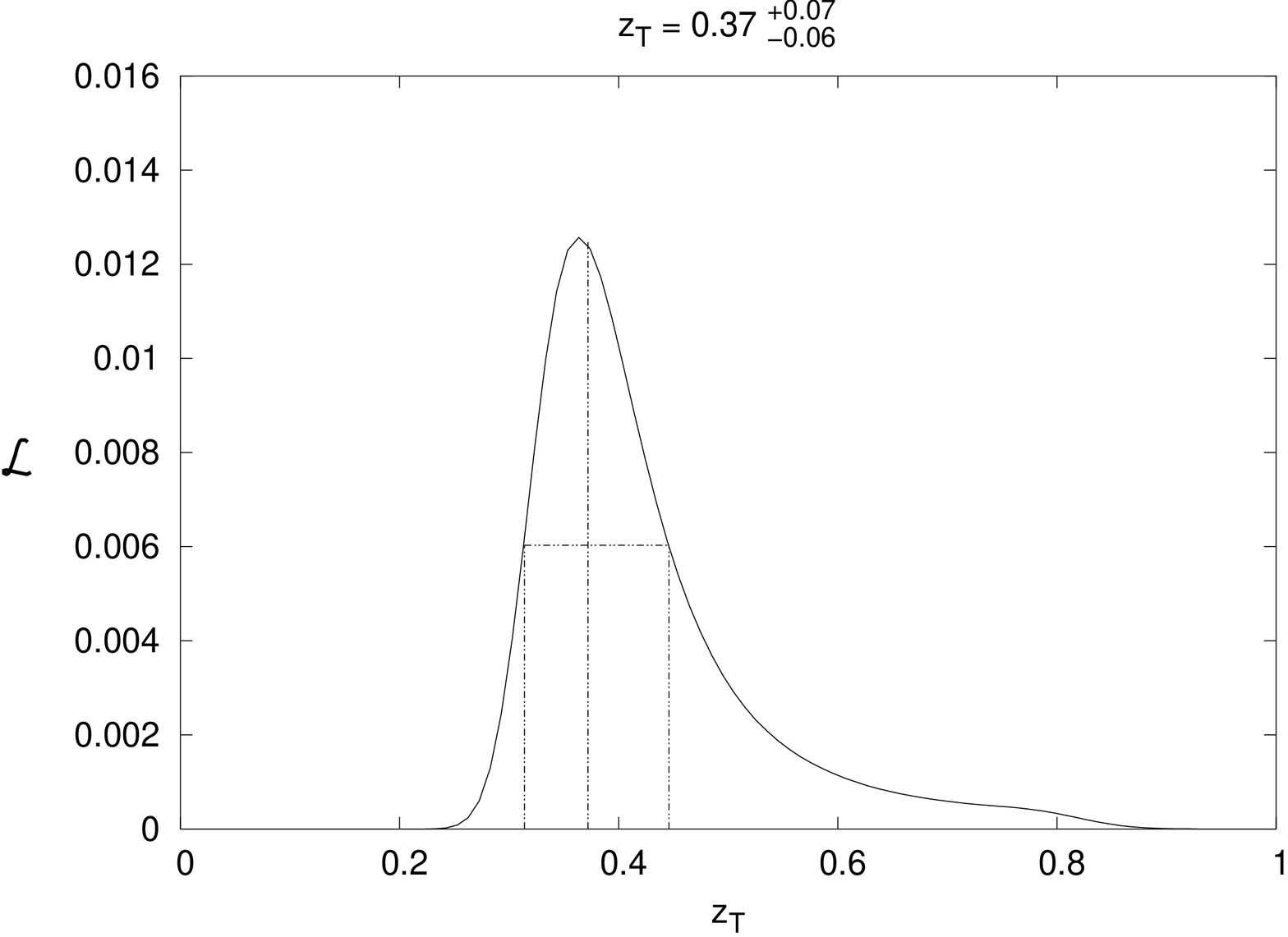} \\ [0.4cm]
\multicolumn{1}{l}{\mbox{\bf (e)}} & 
\multicolumn{1}{l}{\mbox{\bf (f)}} \\ [0.0cm] 
\includegraphics[scale=0.3, angle=270]{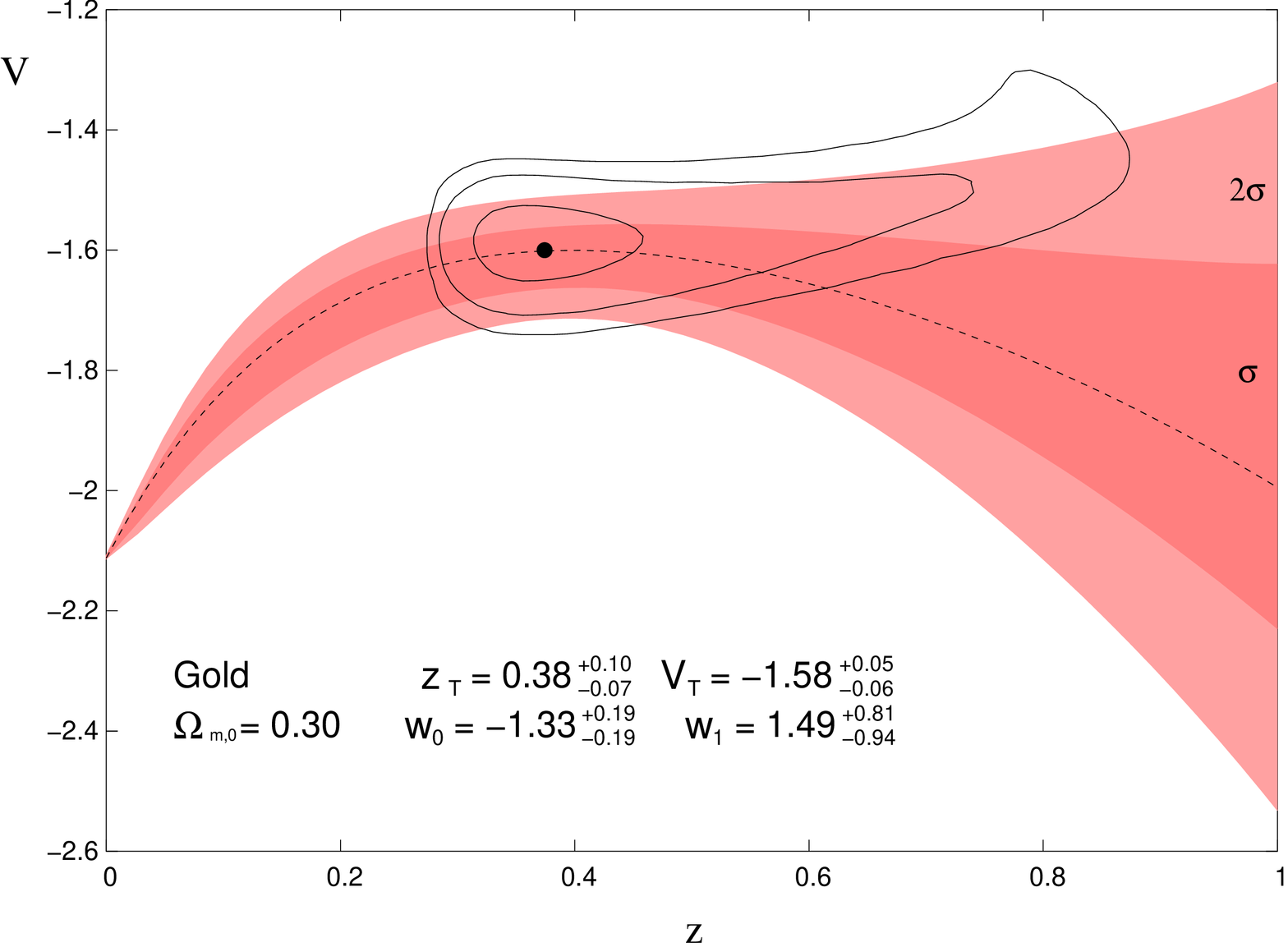} & 
\includegraphics[scale=0.3, angle=270]{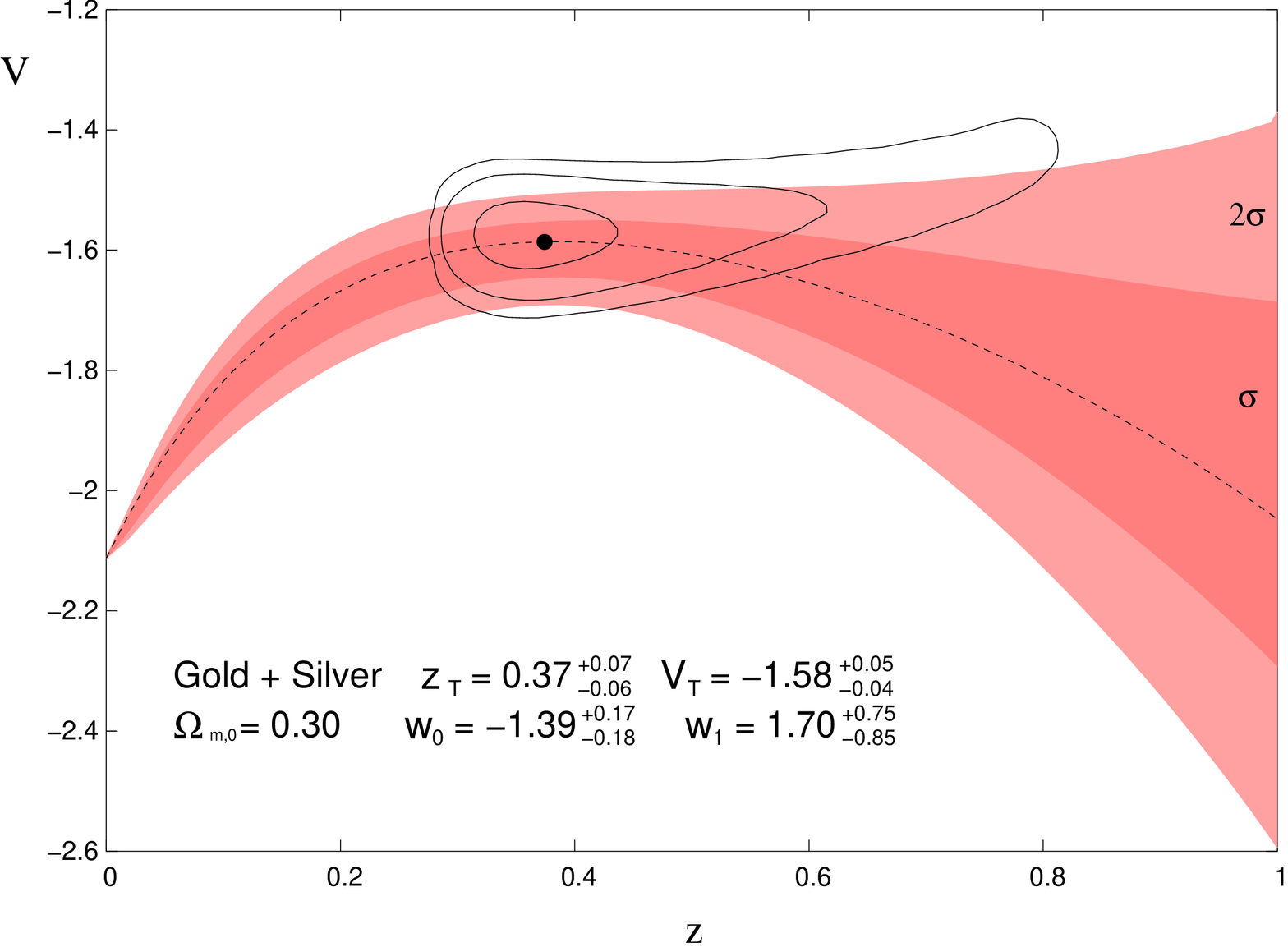} \\ [0.4cm]
\end{array}$
\end{center}
\caption{
The left column refers to the gold subset of Riess et al.'s SN Ia sample 
and the right column to the gold+silver sample. It is assumed that 
$\Omega_{\text{m},0} = 0.30$ and $H_{0} = 65$ ${\rm km}$ ${\rm s^{-1}}$ ${\rm Mpc}$. Figures (a) 
and (b) present the levels of constant transition redshift on the plane $(w_{0},w_{1})$ as well as the 
confidence levels (68\%, 95\% and 99\%) for pairs of $(w_{0}$, $w_{1})$ parameters obtained from the 
maximum likelihood method. The star denotes the best fit values for $w_{0}=-1.35$, $w_{1}=1.52$ 
(gold subset) and for $w_{0}=-1.41$, $w_{1}=1.73$ (gold+silver sample). The dot denotes the cosmological 
constant model. The shaded region is the 1$\sigma$ confidence level for $z_{\text{T}}$ from the likelihood
method around the best fit value (the dashed line). Figures (c) and (d) present the probability 
distribution of the transition redshift $z_{T}$. Figures (e) and (f) show the confidence levels for the 
reconstructed potential function $V(z)$ as well as for the maximum of the potential function 
$(z_{T},V(z_{T}))$. 
}
\label{fig:1}
\end{figure}
\begin{figure}[!ht]
\begin{center}
\includegraphics[scale=0.4, angle=0]{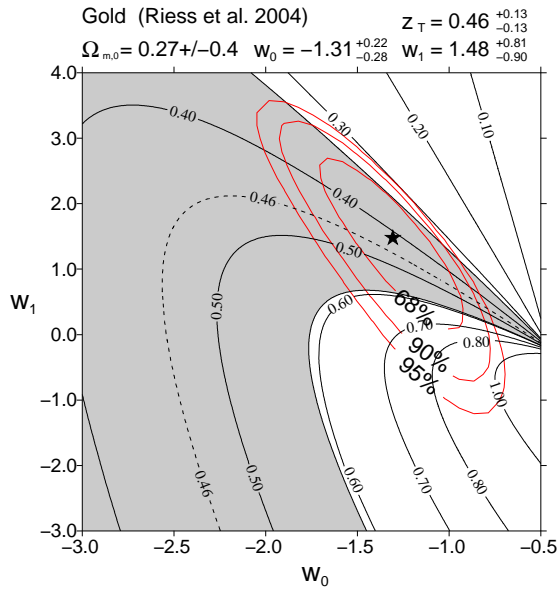} 
\end{center}
\caption{
We draw the levels of constant $z_T$ on the plane $(w_{0},w_{1})$. We distiguish $z_T = 0.46$ (the dashed 
line) with its 1$\sigma$ confidence level. We also mark with the star the best fit of $(w_{0},w_{1})$ and 
the 68\%, 95\% and 99\% confidence levels obtained by Riess et al. \cite{riess04}. The key observation is 
that the transition redshift for their $(w_{0},w_{1})$ parameters is lower than the corresponding value 
obtained from the Turner-Riess kinematic description \cite{turner02}.}
\label{fig:2}
\end{figure}

The best fitted parameters of the model as well as the transition redshift $z_{T}$ and the maximum of the 
potential $V_{T}=V(z_{T})$ are presented in Table~\ref{results1}. 

Assuming that the measurement uncertainties are Gaussian we can simply calculate the likelihood function
$\mathcal{L}(\Omega_{m,0},H_{0},w_{0},w_{1})$ from a chi-squared statistics in the following way
\begin{gather}
\mathcal{L}(\Omega_{m,0},H_{0},w_{0},w_{1})=\frac{1}{N}\exp{\left(-\frac{\chi^{2}(\Omega_{m,0},H_{0},w_{0},w_{1})}{2}\right)},\nonumber \\
N=\iiiint_{-\infty}^{\infty} \exp{\left(-\frac{\chi^{2}(\Omega_{m,0},H_{0},w_{0},w_{1})}{2}\right)}d\Omega_{m,0}dH_{0}dw_{0}dw_{1},
\label{eq:15}
\end{gather}
where $N$ is the normalization coefficient given as an integral of the likelihood function over all 
probability variables.
The confidence levels ($68\%$, $95\%$, $99\%$) for pairs ($w_{0}$,$w_{1}$), obtained for the gold subset 
and full sample, are drawn in Fig.~\ref{fig:1}(a) and Fig.~\ref{fig:1}(b) respectively. We find the 
one-dimensional probability distribution functions for $w_{1}$ and $w_{2}$ separately by integrating the 
likelihood function $\mathcal{L}$ over remaining probability variables and we obtain 
$w_{0}=-1.33^{+0.19}_{-0.19}$, $w_{1}=1.49^{+0.81}_{-0.94}$ for the gold subset and 
$w_{0}=-1.39^{+0.17}_{-0.18}$, $w_{1}=1.70^{+0.75}_{-0.85}$ for the full sample.

We find the probability distribution for the transition redshift $z_{T}$ by integrating the 
likelihood function $\mathcal{L}$ along the levels of constant $z_{T}$ (see Fig.~\ref{fig:1}(c),(d)).
As a result we obtain $z_{T} = 0.38_{-0.07}^{+0.10}$ (gold subset) and $z_{T}= 0.37_{-0.06}^{+0.07}$ 
(gold+silver sample).

We also calculate the probability distribution for the value of the potential $V(z_{T})$ by integrating the 
likelihood function $\mathcal{L}$ along the levels of constant $V(z_{T})$ for the best-fitted value 
of $z_{T}$. In this case we have $V(z_{T}) = -1.58_{-0.06}^{+0.05}$ in units of $10^{3}H_{0}^{2}$ (gold 
subset), $V(z_{T}) = -1.58_{-0.04}^{+0.05}$ in units of $10^{3}H_{0}^{2}$ (gold+silver sample).

Because of the Hamiltonian constraint we can also calculate the value of the Hubble parameter 
corresponding to transition redshift $z_{T}$
\begin{equation}
H_{T} = (1+z_{T}) \sqrt{-2V(z_{T})} 
\end{equation}
and the velocity of the scale factor during the transition epoch
\begin{equation}
\dot{a}_{T} = \sqrt{-2V(z_{T})} 
\end{equation}
and obtain that $H_{T} \simeq 79.2 > H_{0}$ and $\dot{a}_{T} \simeq 56.6 < \dot{a}_{0} = H_{0}$ 
(gold subset). 

The quality of the fitting of the potential function as well as the position of its maximum is 
illustrated in Fig.~\ref{fig:1} (e), (f). It is interesting that the value of $V(z_{T})$ is 
calculated with lower error than $z_{T}$. We also see that the errors decrease in the neighbourhood of 
the region around the maximum of $V(z_{T})$. For the gold+silver sample the errors are lower than 
for the gold subset. 

We estimated the value of $z_{T}$ in the model with linear form of the equation of state coefficient $w(z)$.
The value of transition redshift was also obtained by Riess et al. in the 
different model. To compare their result in context of the presented framework we look for 
the transition redshift for their estimated model (model with linear ansatz on $w(z)$). 
We obtained lower value (marked as a star in Fig.~\ref{fig:2}) than their corresponding value $0.46$ 
(marked as the dashed line). 
In our opinion the reason of this difference is that their estimation of $z_{T}$ was done in different model
without any priors on model parameters.

\section{Conclusions}

We presented the V-method of reconstruction of the potential function. We applied it to the 
estimation of the state parameters of the cosmological model with dark energy. We concentrated 
on the estimation of the position of the maximum of the potential function which can be interpretated as 
the moment of switch from the deceleration to the acceleration phase during the Universe evolution. 

The shape of the potential function is reconstructed from the distant SN Ia data. For our analysis 
we used the samples of SN Ia prepared by Riess et al. 

The Hamiltonian formulation of the dynamics enables us to determine whole dynamics of the 
cosmological model from the potential function only. It is interesting that this function 
determines not only the qualitative structure of the phase space \cite{szydlowski03,szydlowski03c} 
but also gives us some quantitative information about the model parameters. 

We showed that the V-method is very effective in analysis of cosmological models with dark energy 
where we can find the moment of transition as the quantitave parameter. The other obtained parameter of 
state of the system is the velocity of the scale factor (or the Hubble parameter) at the moment of 
transition.  
We can treat these two estimated parameters as initial conditions and determine uniquely the evolution 
path of the model. 

The main results of using this method to the cosmological model with dark energy are  

1. The method allows to study quantitative and qualitative aspects of dynamics of models.

2. We showed that the state parameters at the moment of transition can be obtained in any models for which 
the reconstruction can be done. 

3. We estimated the value of the transition redshift and the value of Hubble function at the transition 
moment moment. 

Riess et al. estimated the model parameters assuming the different forms of the equation of state 
coefficient $w(z)$, but to find the transition redshift $z_{T}$ they used a certain kinematic model with a 
linear expansion for $q(z)$. It seems to be no a priori reasons to extrapolate the value of $z_{T}$ between 
two different models. The question is to which $w(z)$ model corresponds their value of $z_{T}$. 

Our methods overcomes this obstacle. To obtain the model parameters and the transition redshift we use the 
same $d_{L}(z)$ relation.

\begin{acknowledgments}
The paper was supported by KBN grant no. 2 P03D 003 26.
\end{acknowledgments}


\begin{thebibliography}{8}
\expandafter\ifx\csname natexlab\endcsname\relax\def\natexlab#1{#1}\fi
\expandafter\ifx\csname bibnamefont\endcsname\relax
  \def\bibnamefont#1{#1}\fi
\expandafter\ifx\csname bibfnamefont\endcsname\relax
  \def\bibfnamefont#1{#1}\fi
\expandafter\ifx\csname citenamefont\endcsname\relax
  \def\citenamefont#1{#1}\fi
\expandafter\ifx\csname url\endcsname\relax
  \def\url#1{\texttt{#1}}\fi
\expandafter\ifx\csname urlprefix\endcsname\relax\def\urlprefix{URL }\fi
\providecommand{\bibinfo}[2]{#2}
\providecommand{\eprint}[2][]{\url{#2}}

\bibitem[{\citenamefont{Riess et~al.}(1998)}]{riess98}
\bibinfo{author}{\bibfnamefont{A.~G.} \bibnamefont{Riess}}
  \bibnamefont{et~al.}, \bibinfo{journal}{Astron. J.}
  \textbf{\bibinfo{volume}{116}}, \bibinfo{pages}{1009} (\bibinfo{year}{1998}),
  \eprint{astro-ph/9805201}.

\bibitem[{\citenamefont{Perlmutter et~al.}(1999)}]{perlmutter99}
\bibinfo{author}{\bibfnamefont{S.~J.} \bibnamefont{Perlmutter}}
  \bibnamefont{et~al.}, \bibinfo{journal}{Astrophys. J.}
  \textbf{\bibinfo{volume}{517}}, \bibinfo{pages}{565} (\bibinfo{year}{1999}),
  \eprint{stro-ph/9812133}.

\bibitem[{\citenamefont{Riess et~al.}(2004)}]{riess04}
\bibinfo{author}{\bibfnamefont{A.~G.} \bibnamefont{Riess}} \bibnamefont{et~al.}
  (\bibinfo{year}{2004}), \eprint{astro-ph/0402512}.

\bibitem[{\citenamefont{Szydlowski and
  Czaja}(2004{\natexlab{a}})}]{szydlowski03}
\bibinfo{author}{\bibfnamefont{M.}~\bibnamefont{Szydlowski}} \bibnamefont{and}
  \bibinfo{author}{\bibfnamefont{W.}~\bibnamefont{Czaja}},
  \bibinfo{journal}{Phys. Rev. D} \textbf{\bibinfo{volume}{69}},
  \bibinfo{pages}{083518} (\bibinfo{year}{2004}{\natexlab{a}}),
  \eprint{gr-qc/0305033}.

\bibitem[{\citenamefont{Szydlowski and
  Czaja}(2004{\natexlab{b}})}]{szydlowski03c}
\bibinfo{author}{\bibfnamefont{M.}~\bibnamefont{Szydlowski}} \bibnamefont{and}
  \bibinfo{author}{\bibfnamefont{W.}~\bibnamefont{Czaja}},
  \bibinfo{journal}{Phys. Rev. D} \textbf{\bibinfo{volume}{69}},
  \bibinfo{pages}{083507} (\bibinfo{year}{2004}{\natexlab{b}}),
  \eprint{astro-ph/0309191}.

\bibitem[{\citenamefont{Maor et~al.}(2001)\citenamefont{Maor, Brustein, and
  Steinhardt}}]{maor01}
\bibinfo{author}{\bibfnamefont{I.}~\bibnamefont{Maor}},
  \bibinfo{author}{\bibfnamefont{R.}~\bibnamefont{Brustein}}, \bibnamefont{and}
  \bibinfo{author}{\bibfnamefont{P.~J.} \bibnamefont{Steinhardt}},
  \bibinfo{journal}{Phys. Rev. Lett.} \textbf{\bibinfo{volume}{86}},
  \bibinfo{pages}{6} (\bibinfo{year}{2001}).

\bibitem[{\citenamefont{Szydlowski and
  Czaja}(2004{\natexlab{c}})}]{szydlowski04}
\bibinfo{author}{\bibfnamefont{M.}~\bibnamefont{Szydlowski}} \bibnamefont{and}
  \bibinfo{author}{\bibfnamefont{W.}~\bibnamefont{Czaja}}
  (\bibinfo{year}{2004}{\natexlab{c}}), \eprint{astro-ph/0402510}.

\bibitem[{\citenamefont{Turner and Riess}(2002)}]{turner02}
\bibinfo{author}{\bibfnamefont{M.~S.} \bibnamefont{Turner}} \bibnamefont{and}
  \bibinfo{author}{\bibfnamefont{A.~G.} \bibnamefont{Riess}},
  \bibinfo{journal}{Astrophys. J.} \textbf{\bibinfo{volume}{569}},
  \bibinfo{pages}{18} (\bibinfo{year}{2002}), \eprint{astro-ph/0106051}.

\end{thebibliography}
\end{document}